\documentclass[titlepage,12pt]{article}
\usepackage{multirow}
\usepackage{graphicx}
\usepackage{amsmath}
\usepackage{textcomp}
\usepackage{pifont}
\usepackage{blindtext}
\usepackage{authblk}
\usepackage[colorlinks]{hyperref}
\hypersetup{linkcolor=blue,citecolor=black}
\usepackage{ulem}

\DeclareUnicodeCharacter{202F}{\,}

\begin{document}

\title{Sequential tilting 4D-STEM for improved momentum-resolved STEM field mapping}

\renewcommand{\thefootnote}{\fnsymbol{footnote}}
\author[1,2,3]{Christoph Flathmann\thanks{Corresponding author: christoph.flathmann@ruhr-uni-bochum.de}}
\author[1]{Ulrich Ross}
\author[4]{Jürgen Belz}
\author[4]{Andreas Beyer}
\author[4]{Kerstin Volz}
\author[1]{Michael Seibt}
\author[5]{Tobias Meyer\thanks{Corresponding author: tmeyer@uni-goettingen.de}}

\date{May 29, 2025}

\affil[1]{4th Institute of Physics \textendash{} Solids and Nanostructures, University of Goettingen, Friedrich-Hund-Platz 1, 37077 G\"ottingen, Germany}
\affil[2]{Research Center Future Energy Materials and Systems, Ruhr University Bochum, Universitätsstr. 150, 44801 Bochum, Germany}
\affil[3]{Faculty of Physics and Astronomy, Ruhr University Bochum, Universitätsstr. 150, 44801 Bochum, Germany}
\affil[4]{Department of Physics and Materials Science Center, Philipps-University Marburg, Hans-Meerwein Str. 6, 35032 Marburg, Germany}
\affil[5]{Institute of Materials Physics, University of Goettingen, Friedrich-Hund-Platz 1, 37077 G\"ottingen, Germany}

\maketitle
\renewcommand{\thefootnote}{\arabic{footnote}}

\begin{abstract}
Momentum-resolved scanning transmission electron microscopy (MRSTEM) is a powerful phase-contrast technique that can map lateral magnetic and electric fields ranging from the micrometer to the subatomic scale. Resolving fields ranging from a few nanometers to a few hundred nanometers, as well as across material junctions, is particularly important since these fields often determine the functional properties of devices. However, it is also challenging since they are orders of magnitude smaller than atomic electric fields. Thus, subtle changes in diffraction conditions lead to significant changes in the measured MRSTEM signal. One established approach to partially overcome this problem is precession electron diffraction, in which the incident electron beam is continuously precessed while precession-averaged diffraction patterns are acquired. Here, we present an alternative approach in which we sequentially tilt the incident electron beam and record a full diffraction pattern for each tilt and spatial position. This approach requires no hardware modification of the instrument and enables the use of arbitrary beam tilt patterns that can be optimized for specific applications. Furthermore, recording diffraction patterns for every beam tilt allows access to additional information. In this work, we use this information to create virtual large-angle convergent beam electron diffraction (vLACBED) patterns to assess MRSTEM data quality and improve field measurements by applying different data analysis methods beyond simple averaging. The presented data acquisition concept can readily be applied to other 4D-STEM applications.
\end{abstract}

\section{Introduction}

Nanometer-sized electric fields across junctions play a crucial role in electronic devices such as transistors \cite{baliga2010fundamentals}, solar cells \cite{wurfel2016physics} or batteries \cite{yamamoto2010dynamic,haruyama2014space,wang2020situ}, as the associated built-in potentials determine their functional properties. New concepts that improve the performance of these devices require miniaturization and increasingly complex designs. Thus, accurate measurements of the potentials on the relevant length scale of a few to about 100~nm are necessary for technological progress.\par
In recent years, phase-contrast transmission electron microscopy (TEM) techniques have demonstrated great potential for accomplishing this task. One such method with a long tradition of mapping potentials across junctions is off-axis electron holography \cite{tonomura1987applications,lichte2007electron}. It has been used to study silicon p-n junctions \cite{frabboni1985electron,rau1999two,twitchett2002quantitative,cooper2016off}, compound semiconductors \cite{sasaki2014direct,anada2019accurate} and quantum wells \cite{cooper2021mapping,cooper2023mapping}. However, off-axis electron holography has several drawbacks, such as the requirement of an undistorted reference wave and a limited field of view.\par
An alternative approach to mapping fields involves measuring the deflection of the transmitted beam in scanning transmission electron microscopy (STEM) mode. In this case, one measures the average lateral momentum transferred from the sample to the electron beam via the center of mass (COM) of the diffraction pattern \cite{muller2014atomic}, which is related to the lateral gradient of the projected electrostatic potential convolved with the incident probe intensity \cite{winkler2020direct}.\par
There are two methods for measuring the average momentum transfer: differential phase contrast (DPC) STEM and momentum-resolved STEM (MRSTEM). DPC uses a segmented detector to approximate the COM, allowing for the rapid acquisition of data, but is rather sensitive to the alignment of the experimental setup. Nevertheless, DPC has been used to map fields across p-n junctions and polarization fields in quantum wells \cite{shibata2015imaging,seki2017quantitative,clark2018probing,haas2019direct,toyama2020quantitative}. With the advent of faster pixelated detectors, MRSTEM has become more common. Here, a full diffraction pattern is acquired at each scan position, enabling more robust data acquisition and post-processing than DPC. MRSTEM has been used to measure fields across p-n homojunctions with and without an applied bias, as well as potential drops across heterointerfaces \cite{bruas2020improved,beyer2021quantitative,da2022assessment,chejarla2023measuring}.\par
However, all the phase-contrast TEM methods have issues with quantitatively and reliably measuring extended fields. This is because fields in the nanometer range typically have absolute values below 0.1 V/nm, which is orders of magnitude smaller than the gradients of atomic electric potentials. Thus, even small changes in the often dynamical diffraction conditions across the field of view lead to significant changes in the diffraction intensity, resulting in large and hard-to-predict signals in phase contrast TEM. This is particularly pronounced in complex samples, such as heterostructures, where strain gradients, defects, thickness variations, different lattice constants or symmetries, and local misorientation can strongly distort field measurements making it challenging to achieve accurate measurements in real devices \cite{mahr2022towards,heimes2023impact}.\par
The common approach to mitigate this issue involves combining DPC or MRSTEM with precession \cite{bruas2020improved,chejarla2023measuring,mawson2020suppressing}. In this method, the incident electron beam continuously precesses around the optical axis at a polar angle of typically around 10~mrad. The precession is hardware-synchronized with the scan, and the diffraction signal is averaged over all azimuthal angles. This incoherent average of diffraction patterns has been shown to suppress dynamical diffraction effects as compared to individual diffraction patterns \cite{midgley2015precession}, thus significantly improving the results of DPC and MRSTEM \cite{chejarla2023measuring}.
A comparison of the strengths and weaknesses of these convergent beam methods compared to off-axis electron holography using a plane incident wave can be found in reference \cite{cooper2024measuring} with the conclusion that for a proper crystal orientation a plane wave is beneficial since convergent beams more likely include incident directions with strong dynamical effects.
\par 
The approach presented here goes beyond standard precession electron diffraction and offers the possibility to reduce these dynamical diffraction effects: It sequentially records full diffraction patterns for each scan position and tilt of the incident electron beam. This provides 6-dimensional data sets, which offer more information than previous methods. It also gives us the ability to use arbitrary tilt patterns, which has been reported to be beneficial for MRSTEM \cite{mawson2020suppressing}. The underlying principle of this sequential tilting MRSTEM is very similar to the tilt scan averaged DPC STEM presented in references \cite{murakami2020magnetic,kohno2022development,toyama2022quantitative,toyama2023real,toyama2025nanoscale}, except that we have access to the full 6D data.\par
In order to experimentally verify our approache, we use sequential tilting MRSTEM on \mbox{AlGaAs/GaAs} heterojunctions, a well-studied model system. We use the additional information contained in the sequential tilted MRSTEM data to assess the data quality, post-process the data for different sample orientations to reduce dynamical diffraction effects, and confirm that this additional information is essential to improving the reliability and quantifiability of MRSTEM results.

\section{Experimental}

\begin{figure*}
    \centering
    \includegraphics[width=0.8\textwidth]{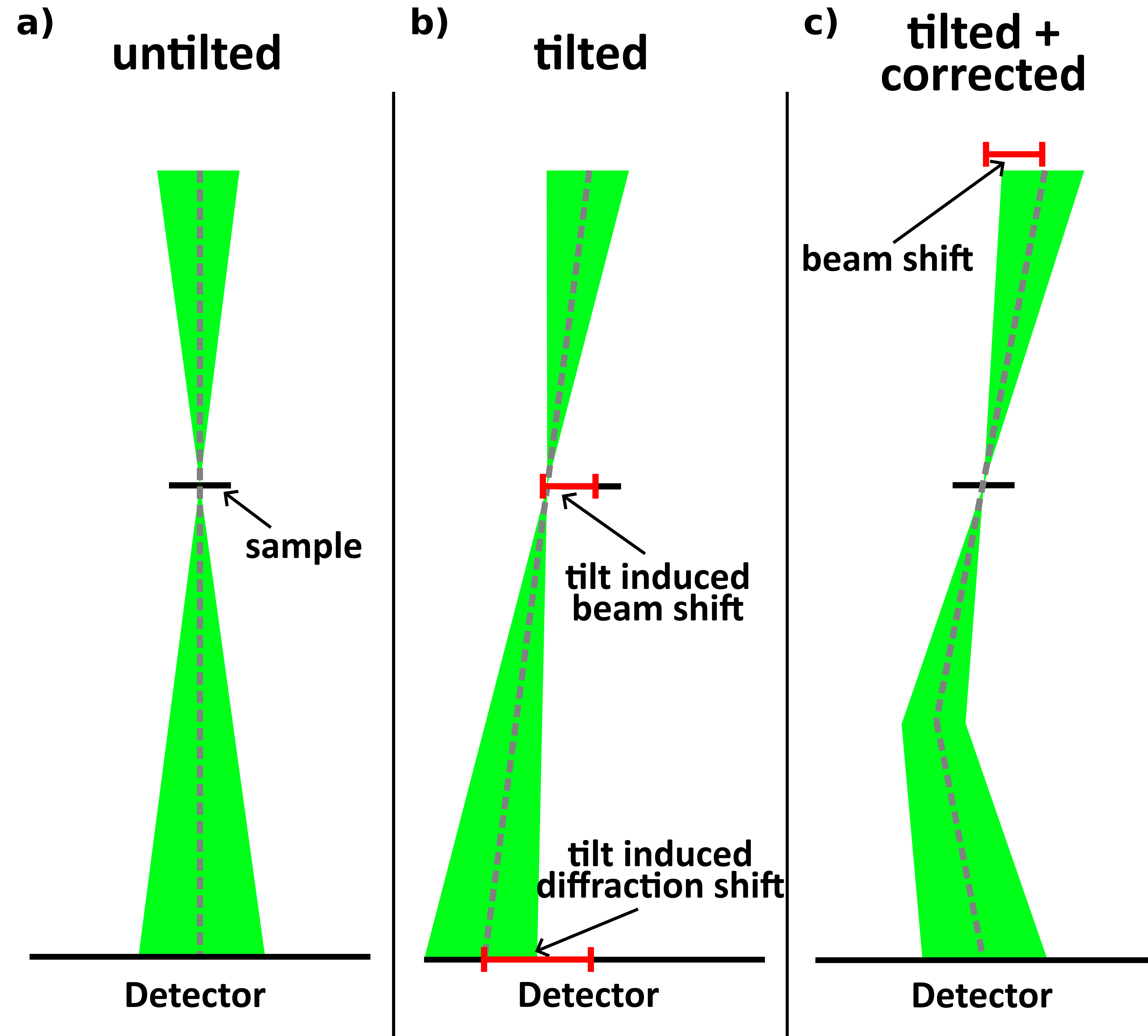}
    \caption{
    Beam tilt and detilt: \textbf{a)} shows the beam path without tilt. \textbf{b)} shows the beam path with tilt. \textbf{c)} shows the beam path with tilt, detilt and tilt induced shift correction. The relevant shifts are highlighted in the images.}
    \label{tilt_overview}
\end{figure*}

\subsection{Calibration procedure}
To achieve precise beam tilts without hardware modifications, we implemented a custom-developed calibration and acquisition procedure. A two-step calibration process is necessary to enable the sequential tilting MRSTEM: first, calibrate the beam tilts; then, calibrate the tilt-induced beam shift. Figure~\ref{tilt_overview}~\textbf{a)}-\textbf{c)} shows a schematic of the beam paths for different tilt and calibration conditions, with visualizations of the relevant parameters.\par
The first step is to calibrate the microscope's tilt coils via cross-correlation of diffraction patterns acquired for different beam tilts in vacuum. The calibration script accomplishes this by addressing the tilt coils with the TEM scripting interface command. However, since the input values of the command are in arbitrary units, a very small value is initially entered and gradually increased until the shift of the diffraction pattern on the detector reaches a threshold of at least 1/4 of the detector. 
For an known pixel spacing of the detector, the shift of the diffraction pattern provides a first quantitative beam tilt calibration which is then used as input for the fine calibration.\par
In the fine calibration, the beam is tilted along a circular pattern with a user-defined tilt angle (based on the estimated calibration from the previous step) and a specified number of samples. By measuring the shift of the diffraction pattern for each beam tilt, one obtains several pairs of desired and actual tilt angles of the beam. Assuming a linear relationship between the desired and actual beam tilts, the resulting system of linear equations can be solved to calibrate the beam tilt coils. This calibration process is carried out for the beam tilt coils above the sample and the detilt (diffraction shift) coils below the sample. This gives us the capability to tilt the beam above the sample while keeping the diffraction pattern stationary on the detector.\par
The second step is to calibrate the tilt-induced beam shift by cross-correlating STEM images. To measure this shift, the beam is tilted again along a circular pattern with a user-defined tilt angle and number of samples. This captures the tilt-induced beam shift in a similar way as the precession path segmentation, described in reference~\cite{nordahl2023correcting}, used for scanning precession electron diffraction. The beam shift is measured via cross-correlation of the STEM images, and the calibration is obtained in the same manner as for the tilt coil calibration. Next, the calibration script addresses the beam shift coils of the microscope, and the corresponding beam shifts are measured via cross-correlation of the images. Similar to calibrating the tilt coils, this process involves first a coarse calibration, followed by a fine calibration. Calibrating the tilt-induced beam shift is usually more challenging than calibrating the beam tilt coils since image contrast changes with beam tilt.

\subsection{Samples}
In this study, we examine the potential distribution across heterojunctions of a nominally undoped 10-fold \mbox{Al$_{0.3}$Ga$_{0.7}$As/GaAs} super lattice sample. The sample was grown via metal-organic vapor phase epitaxy (MOVPE) in an AIX 200 GFR reactor on semi insulating GaAs(001) substrate. The actual Al content was determined from high-resoltuion X-ray diffraction by fitting a 004 \mbox{$\omega$/2$\theta$-scan}. It is an excellent model system because the expected MRSTEM signal results almost exclusively from a change in the mean inner potential at the interface of the two materials. Therefore, we can accurately estimate the potential profile across the \mbox{AlGaAs/GaAs} interface. Additionally, \mbox{AlGaAs} and \mbox{GaAs} have a negligible lattice mismatch of approximately $\sim$0.3~\% at room temperature, resulting in minimal strain relaxation in the sample. However, the two materials exhibit significant differences in diffraction properties due to their distinct atomic form factors ($f_{Ga}\approx f_{As}\neq f_{Al}$). Therefore, we expect a substantial change in diffraction conditions across the junction but rather constant diffraction conditions on either side of the junction.

\subsection{Equipment \& Measurements}
For this study, we used an image corrected Titan 80–300 G2 ETEM operated at 300~kV, equipped with a Gatan Quantum 965 ER image filter and a Gatan UltraScan 1000XP CCD camera.\par
We prepared cross-section TEM lamellae in [110] orientation from the sample using a standard FIB procedure in a Thermo Fisher Scientific Helios G4. The lamellae thickness was measured as 330 nm by electron energy loss spectroscopy using the log-ratio method, assuming an inelastic mean free path of 145~nm according to the method proposed by Iakoubovskii et al. \cite{iakoubovskii2008thickness}. Using such thick lamellae is beneficial for mapping nanometer-sized electric fields because it reduces the influence of surface effects.\par
\begin{figure*}
    \centering
    \includegraphics[width=0.9\textwidth]{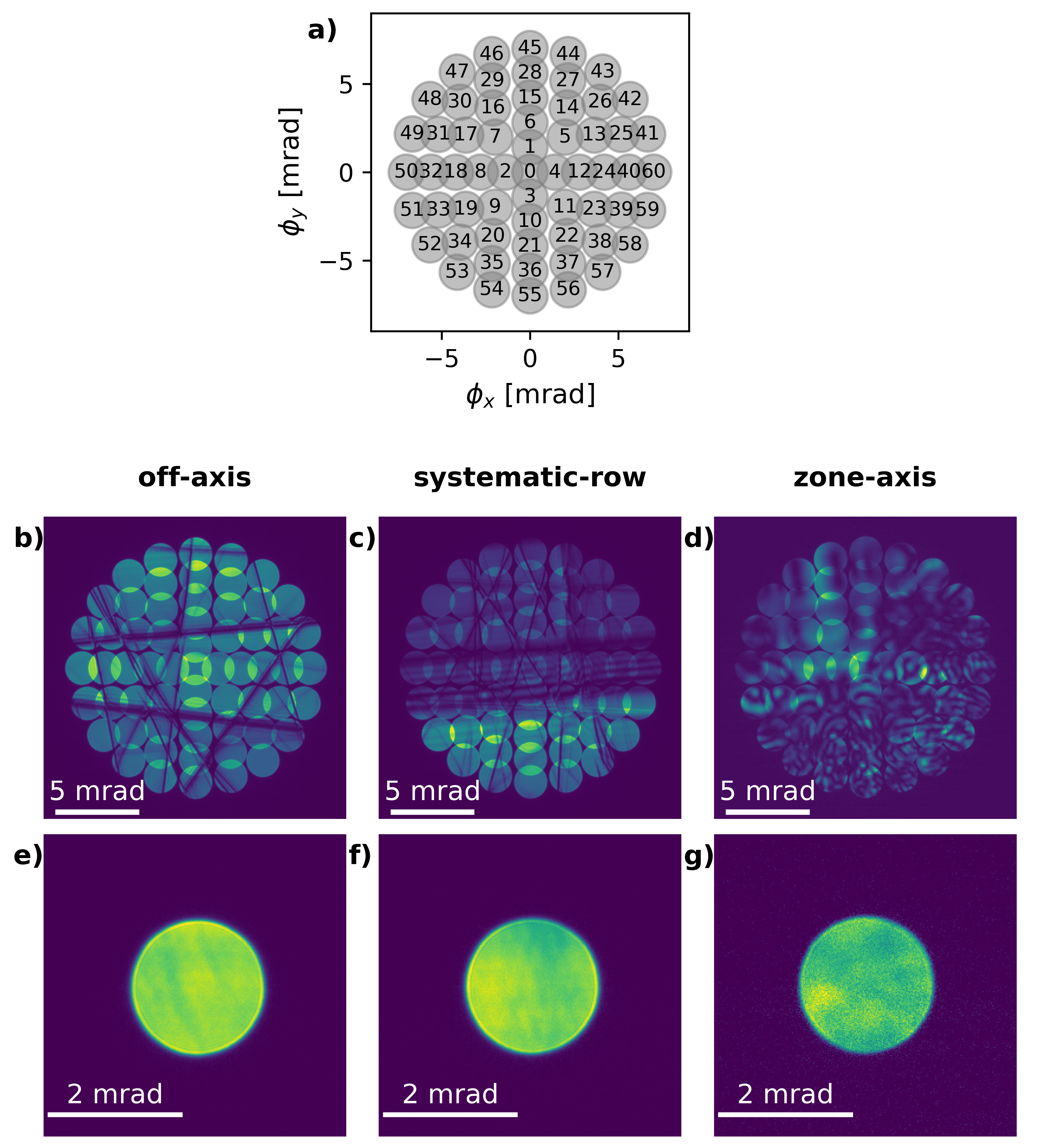}
    \caption{Beam tilt pattern and diffraction patterns: Figure \textbf{a)} shows the beam tilt pattern used throughout this work. Images \textbf{b)}-\textbf{d)} show the resulting vLACBED patterns for a field-free part of the sample on the GaAs side for three different sample orientations. The same diffraction patterns used for creating the vLACBED patterns are shown as averaged diffraction patterns in \textbf{e)}-\textbf{g)}.}
    \label{diffraction_patterns}
\end{figure*}

All of the MRSTEM results presented in this study were acquired with a semiconvergence angle of 1~mrad, resulting in a diffraction-limited probe size of $\sim$2~nm, and a probe current of $\sim$40~pA. Sequential tilting measurements were carried out as line scans across the junction with 30~nm position averaging perpendicular to the scan direction. The beam tilt pattern was always disc-shaped and consisted of 61 beam tilts with a maximum tilt angle of 7~mrad, as shown in Figure~\ref{diffraction_patterns}~\textbf{a)}. Diffraction patterns were acquired using a custom Gatan DigitalMicrograph\textsuperscript{\textregistered} script \cite{meyer2021structural} with 0.1~s exposure time, and the detector was binned to 256$\times$256~pixels, resulting in a pixel size of 17.5~$\mu$rad. This produced diffraction patterns containing only the direct diffraction disc. Furthermore, the diffraction patterns were energy filtered using a 10~eV  slit. To remove artifacts such as descan from the data, a vacuum reference was taken under identical conditions for every scan.

\section{Results \& Discussion}

\subsection{Calibration quality}
To get an initial idea of the tilt calibration quality, we can create a virtual large-angle convergent beam electron diffraction (vLACBED) pattern using the diffraction patterns from individual beam tilts at the same position. This is done by shifting the diffraction patterns according to their respective beam tilt angles. The vLACBED patterns for a GaAs position are shown in Figure~\ref{diffraction_patterns}~\textbf{b)}-\textbf{d)} for three different sample orientations. In particular, for the off-axis orientation, one can see that all higher-order Laue zone (HOLZ) lines (narrow features) and Bragg deficiency lines (broad features) run continuously through the vLACBED pattern, indicating that both tilt and detilt are precisely calibrated.\par
\begin{figure*}
    \centering
    \includegraphics[width=\textwidth]{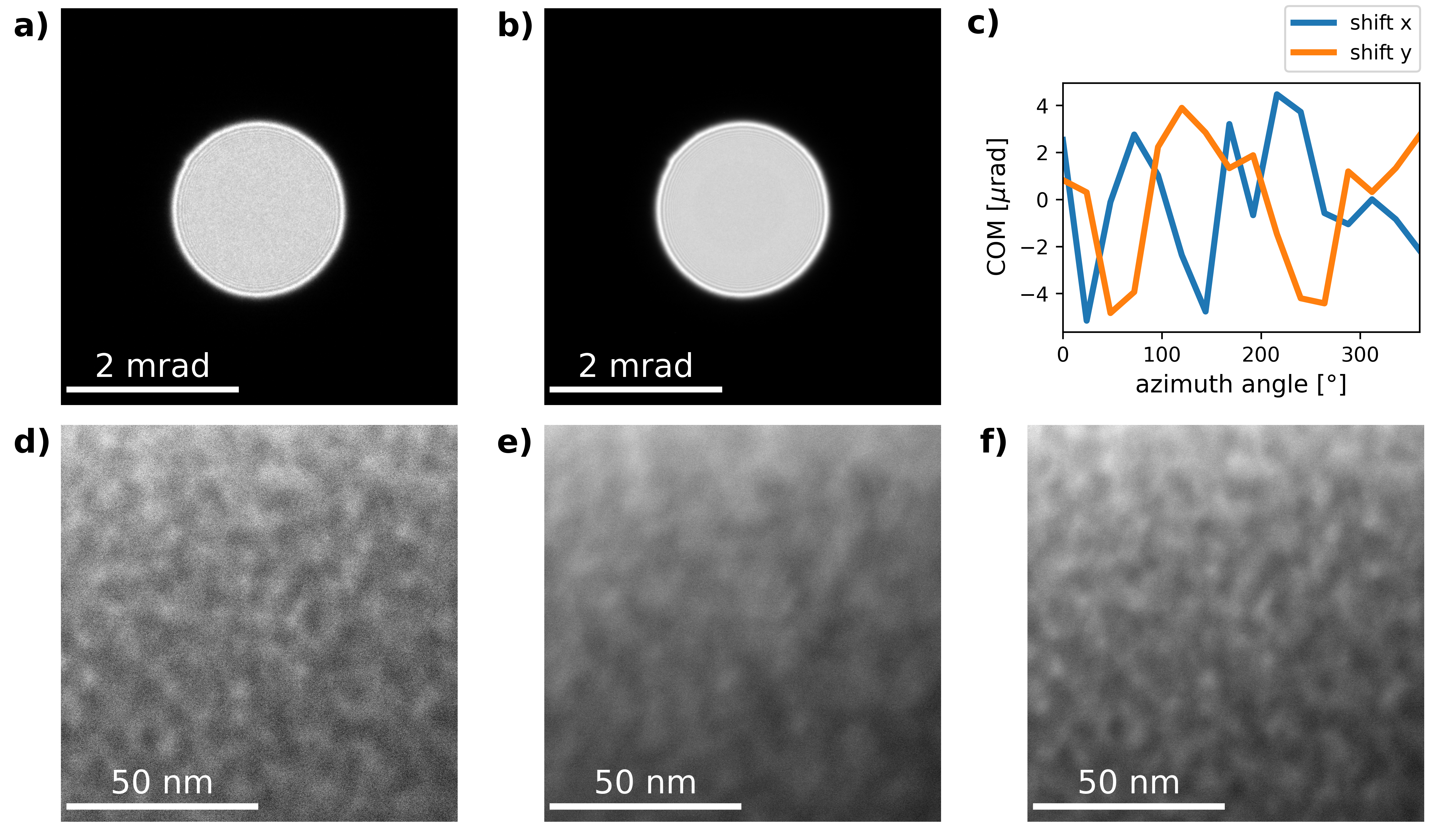}
    \caption{Calibrations of beam tilt and tilt-induced beam shift: \textbf{a)} shows a diffraction pattern acquired in vacuum without beam tilt. \textbf{b)} shows the sum of 16 diffraction patterns tilted along a circle with a radius of 7~mrad, under the same conditions as in \textbf{a)}. \textbf{c)} shows the change in COM of the 16 diffraction patterns of \textbf{b)} relative to the COM of the diffraction pattern of \textbf{a)} as a function of azimuth angle. \textbf{d)} shows an ADF image of FIB-deposited Pt particles acquired without beam tilt. \textbf{e)} shows the sum of 16 ADF images acquired with beam tilts along a circle with a radius of 7~mrad from the same region without tilt-induced beam shift correction. Figure \textbf{f)} shows the same image as \textbf{e)}, but with tilt-induced beam shift correction.}
    \label{calibration}
\end{figure*}

To analyze the tilt-induced diffraction shift remaining after calibration, we use an untilted vacuum diffraction pattern, shown in Figure~\ref{calibration}~\textbf{a)}, and 16 vacuum diffraction patterns tilted along a 7~mrad circular pattern. The sum of these patterns is shown in Figure~\ref{calibration}~\textbf{b)}. A visual assessment of Figure~\ref{calibration}~\textbf{b)} shows no distortion of the diffraction pattern. To quantitatively analyze the remaining tilt-induced diffraction shift, we subtracted the COM of the untilted diffraction pattern from the COMs of the tilted diffraction patterns. The result is plotted as a function of the azimuth angle in Figure~\ref{calibration}~\textbf{c)}, revealing an RMS tilt-induced diffraction shift of $\sim$4~$\mu$rad, corresponding to $\sim$1/4 of a pixel.\par 
The influence of tilt-induced beam shift is analyzed by similar means with an untilted beam and 16 tilted beams as described in the previous paragraph. Figure~\ref{calibration}~\textbf{d)} shows an ADF image of FIB deposited Pt:C acquired without a beam tilt and Figure~\ref{calibration}~\textbf{e)} shows the sum of the ADF images acquired with the 16 tilted beams from the same region without tilt-induced beam shift calibration. Tilting the beam clearly leads to blurring of the summed image. When the same image is created with tilt-induced beam shift correction, as shown in Figure~\ref{calibration}~\textbf{f)}, the image shows much less blur. This is confirmed by a decrease of the RMS tilt-induced beam shift from $\sim$4~nm to $\sim$1~nm, determined by cross-correlation of the ADF images. Thus, after calibration, the resulting RMS tilt-induced beam shift is smaller than the probe size.

\subsection{Evaluating MRSTEM data}

\begin{figure*}
    \centering
    \includegraphics[width=\textwidth]{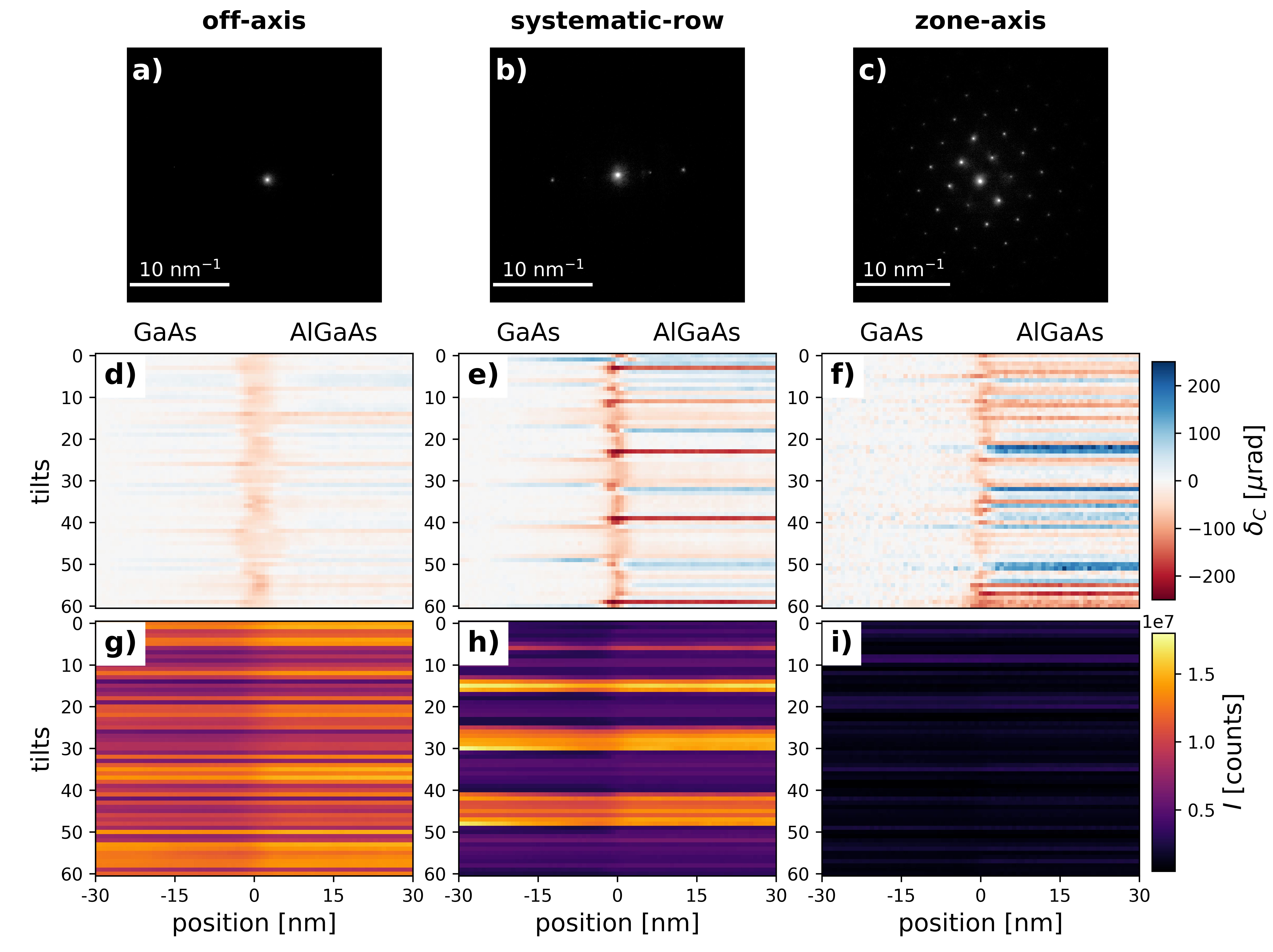}
    \caption{Residual COM and total intensity of the acquired diffraction patterns for different sample orientations: Images \textbf{a)}-\textbf{c)} show the SAED patterns for the off-axis, systematic-row and zone-axis orientation. Images \textbf{d)}-\textbf{f)} show the corresponding COM for each beam tilt and scan position, calculated by subtracting the average COM from a field-free region (-30~nm to -20~nm). Images \textbf{g)}-\textbf{i)} show the total intensity of the acquired diffraction patterns for each beam tilt and scan position. The \mbox{GaAs/AlGaAs} interface is located at 0~nm.}
    \label{COM_shifts}
\end{figure*}

Sequential tilting 4D-STEM provides more opportunities to evaluate the quality of MRSTEM data than conventional methods. This is evident when comparing the vLACBED pattern in Figure~\ref{diffraction_patterns}~\textbf{b)}-\textbf{d)} and the averaged diffraction pattern in Figure~\ref{diffraction_patterns}~\textbf{e)}-\textbf{g)}, which originate from the same data. For instance, the HOLZ lines and Bragg deficiency lines in the vLACBED pattern can be used to determine the sample orientation dependent on the scan position from sequential tilted 4D-STEM data. This information can then be used to optimize MRSTEM data processing.\par
Here, we evaluate the quality of the data for three different sample orientations: off-axis, systematic-row, and zone-axis. The zone-axis orientation refers to a [110] orientation of the lamella. The systematic-row orientation includes an 8.5$^\circ$ rotation around the interface normal with respect to the zone-axis orientation. The off-axis orientation adds another 0.9$^\circ$ rotation parallel to the interface normal. Thus, we expect the off-axis orientation to have the least dynamic diffraction conditions, but also an interface blur of about 5~nm. The corresponding selected area electron diffraction (SAED) patterns are shown in Figure~\ref{COM_shifts}~\textbf{a)}-\textbf{c)}.\par
For each sample orientation, we acquire sequential tilting 4D-STEM line scans across the \mbox{AlGaAs/GaAs} interface. We measure and correct the drift using ADF reference images taken after each sequential scan. From the drift-corrected MRSTEM data, we calculate the COM for every beam tilt and position. To correct the data for descan, we apply the same processing to an MRSTEM vacuum reference and subtract the resulting reference COM from the measurement's COM. Furthermore, we rotate the COM vector to account for the difference in rotation between the sample and detector coordinate systems. For the following steps, we only use the COM perpendicular to the interface, as no field is expected to be parallel to the interface and no COM shift is observed in this direction. By subtracting the mean COM of a field-free region for each beam tilt, we obtain position-dependent COM changes relative to the field-free reference.
\begin{equation}
    \delta_{C}(\pmb{r},\pmb{K})=\langle \pmb{p} \rangle (\pmb{r},\pmb{K})-\overline{\langle \pmb{p} \rangle}(\pmb{K})
\end{equation}
where $\langle \pmb{p} \rangle (\pmb{r},\pmb{K})$ is the COM for a given scan position $\pmb{r}$ and beam tilt $\pmb{K}$ and $\overline{\langle \pmb{p} \rangle}(\pmb{K})$ is the mean COM from a field free region. Thus, the magnitude of $\delta_{C}$ measures how strongly the COM changes, either due to local fields or local changes of the diffraction condition. $\delta_{C}$ is plotted for the three sample orientations in Figure~\ref{COM_shifts}~\textbf{d)}-\textbf{f)}, where $\overline{\langle \pmb{p} \rangle}(\pmb{K})$ was averaged between -20~nm and -30~nm on the \mbox{GaAs} side. As expected, the $\delta_{C}$ maps show small variations on the \mbox{GaAs}  side of the junction and the smallest variations for off-axis orientation. The other two orientations exhibit a significant increase in $\delta_{C}$ on the entire \mbox{AlGaAs} side of the junction, indicating that a change in diffraction conditions causes the COM shift. For the systematic-row orientation, only a few beam tilts show strong COM shifts on the \mbox{AlGaAs} side. For the zone-axis orientation, most beam tilts show strong COM shifts on the \mbox{AlGaAs} side. This suggests that, in the systematic-row orientation, certain beam tilts remain relatively unaffected by changes in diffraction conditions. In contrast, close to the zone-axis, almost all beam tilts are significantly impacted by these changes. We will use this insight to optimize the data processing in the following section.\par
Furthermore, we can plot the total intensity of the acquired diffraction patterns as a function of scan position and beam tilt, i.e., tilt-dependent virtual bright field images, as shown in Figure~\ref{COM_shifts}~\textbf{g)}-\textbf{i)}. When we compare the intensities to the change in the COM across the interface, we observe a correlation between low intensities and significant COM changes for off-axis and systematic-row orientations. This is to be expected, as the acquired diffraction patterns only contain the direct beam, and more dynamic diffraction conditions result in lower intensity in the direct beam for off-axis and systematic-row orientations. For zone-axis orientation, no clear correlation is observed between the COM change across the interface and the intensity.

\subsection{Potential mapping}

\begin{figure*}
    \centering
    \includegraphics[width=\textwidth]{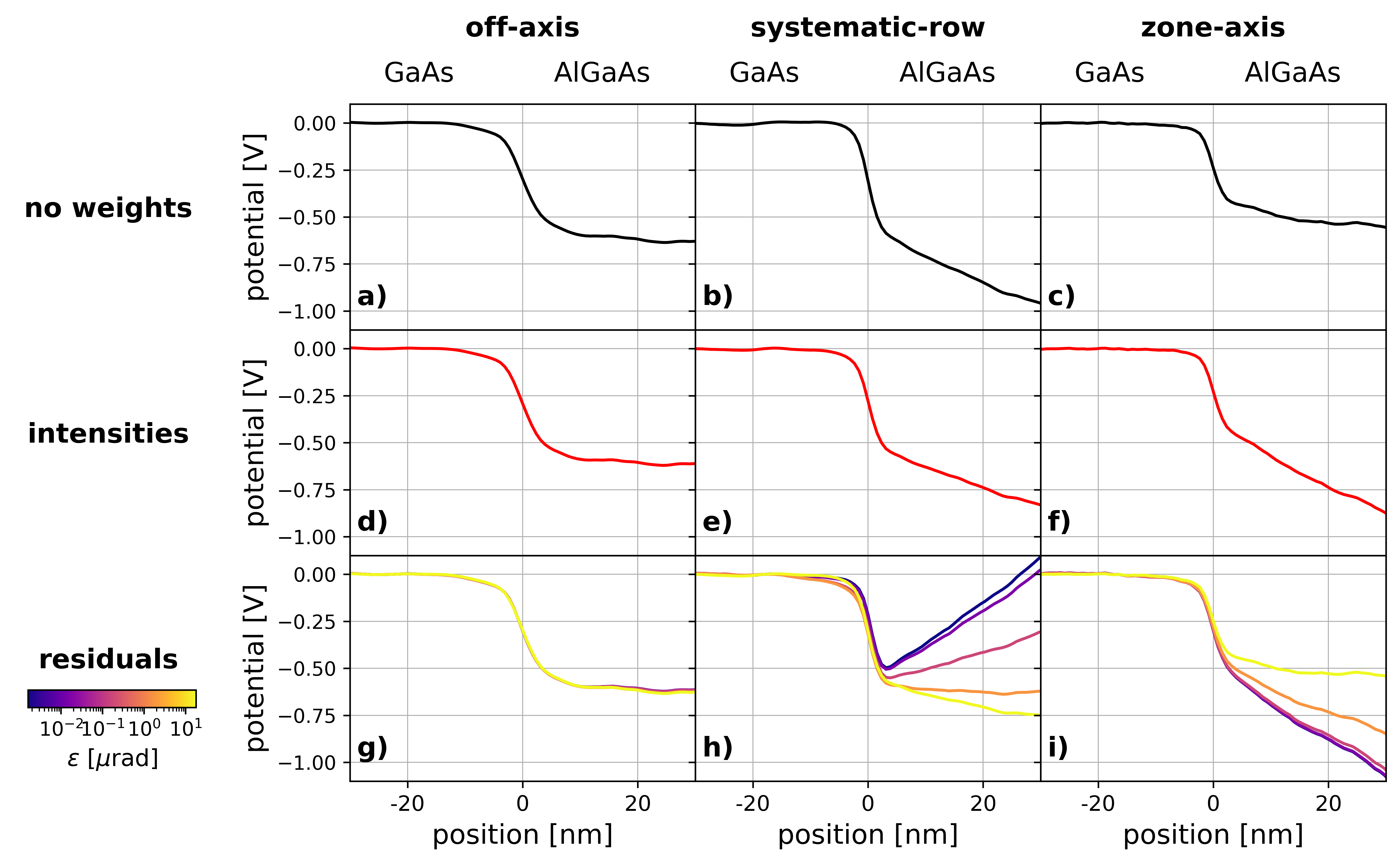}
    \caption{Potential distributions across the AlGaAs/GaAs junction for different sample orientations and post-processing treatments: \textbf{a)}-\textbf{c)} show the potential distributions across the interface for different sample orientations when the COM is first calculated for every beam tilt and then averaged. \textbf{d)}-\textbf{f)} show potential distributions obtained from the same data, but with the COMs weighted by their intensities. \textbf{g}-\textbf{i} show potential distributions obtained using residual COMs for weighting as a function of the regularization parameter $\epsilon$. The \mbox{GaAs/AlGaAs} interface is located at 0~nm.}
    \label{overview_potential}
\end{figure*}

In general, we can calculate the tilt averaged COM
\begin{equation}
    \overline{\langle \pmb{p} \rangle}(\pmb{r})=\frac{\sum_{\pmb{K}=0}^N a(\pmb{r},\pmb{K})\langle \pmb{p} \rangle (\pmb{r},\pmb{K})}{\sum_{\pmb{K}=0}^N a(\pmb{r},\pmb{K})}
\end{equation}
where $a(\pmb{r},\pmb{K})$ is a tilt and position dependent weighting factor.\par
The simplest approach is to set all $a(\pmb{r},\pmb{K})$ to 1, which is different from the conventional precession MRSTEM approach, where the diffraction patterns are first averaged and then the COM is calculated. Thus, coventional precession MRSTEM implicitly weights the tilted diffraction patterns by their intensity. We can reproduce this weighting by setting $a(\pmb{r},\pmb{K})$ to the total intensity of the respective diffraction pattern. Furthermore, we can construct other weights using a priori knowledge about the sample. We can design weights that are high when the COM residuals between field-free parts on opposite sides of the junction are small.
\begin{equation}
    a(\pmb{K})=\frac{1}{|\overline{\langle \pmb{p} \rangle} (\pmb{K})_{GaAs}-\overline{\langle \pmb{p} \rangle} (\pmb{K})_{AlGaAs}|+\epsilon}
\end{equation}
where $\overline{\langle \pmb{p} \rangle} (\pmb{K})_{GaAs}$ and $\overline{\langle \pmb{p} \rangle} (\pmb{K})_{AlGaAs}$ are averaged over field free regions on the \mbox{GaAs} and \mbox{AlGaAs} side respectively and $\epsilon$ is a small regularization constant to avoid diverging weights.\par
We use the resulting tilt-averaged COM values to calculate the electric field in the sample. We do this by using the interaction constant and lamella thickness while assuming the field is constant along the electron beam direction. By integrating the electric field along the scan direction, we obtain the potential distribution across the interface. Figure~\ref{overview_potential} summarizes and compares the potential distributions across the \mbox{AlGaAs/GaAs} interface for different sample orientations and analysis strategies.\par
Beginning the discussion regarding the off-axis orientation shown in Figure~\ref{overview_potential}~\textbf{a)}, \textbf{d)} and \textbf{g)}, we observe an excellent agreement between the measured and expected potentials. The potential drops at the interface and remains constant on both sides of the junction. This is true for all types of applied data analysis. In general, MRSTEM measurements appear quite reliable under these weak diffraction conditions.\par
The situation is different for the systematic-row orientation, shown in Figure~\ref{overview_potential}~\textbf{b)}, \textbf{e)} and \textbf{h)}. In this case, there is a linear decrease in the measured potential on the field-free \mbox{AlGaAs} side without weights, indicating that a change in the diffraction condition causes an offset in the COM. The intensity-weighted potential exhibits slightly better behavior than the potential without weights, which can be understood intuitively by considering the intensity dependence of the direct diffraction disc. In this orientation, stronger diffraction results in more intensity being transferred to the diffracted discs, resulting in less intensity in the direct disc. Thus, intensity is an effective weight for suppressing contributions from beam tilts with strong diffraction under these conditions. Moreover, constructing the weights from the residual COM with the proper choice of $\epsilon$, in this case $\epsilon\approx 1~\mu$rad, yields a significantly improved result. Since the optimally processed potential distribution almost exactly matches the corresponding off-axis potential distributions, it is evident that proper data processing mitigates the impact of dynamic diffraction on the MRSTEM results.\par
The behavior is different for the zone-axis orientation shown in Figure~\ref{overview_potential}~\textbf{c)}, \textbf{f)} and \textbf{i)}. No weighting yields better results than intensity weighting. This is because the correlation between dynamic diffraction artifacts in the MRSTEM data and the intensity of the direct disc is absent under strongly dynamic diffraction conditions. Thus, the improvement by tilt averaging is solely due to incoherently averaging more diffraction patterns. When weights are calculated based on the residual COM, the best result is obtained with a large $\epsilon$ value. This leads to equal weights for all beam tilts, resulting in the same outcome as without weights. This indicates that all beam tilts are strongly affected by dynamic diffraction for this sample orientation, and excluding data from the averaging process only worsens the MRSTEM results.

\section{Conclusion}

In this work, we presented an implementation of sequential tilting 4D-STEM. Our implementation enables robust alignment of beam tilt and detilt, yielding a residual diffraction shift of $\sim$4~$\mu$rad and a tilt-induced beam shift of $\sim$1~nm, without requiring additional hardware. Furthermore, we tested several strategies for using the additional information obtained through sequential tilting 4D-STEM to improve the reliability of MRSTEM results under different diffraction conditions. In doing so, we identified different regimes of dynamic diffraction by accessing MRSTEM data from individual beam tilts. Sequential tilting MRSTEM generally yields good results for weak diffraction conditions. For medium and strong diffraction conditions, we improved the sequential tilting MRSTEM results through appropriate data processing, which are inherently inaccessible for previous MRSTEM techniques. This demonstrates the benefits of sequential tilting 4D-STEM for the reliability and accuracy of MRSTEM measurements.\par
We are convinced that using direct electron detectors, improving beam tilt patterns, and optimizing data post-processing will further improve the results and will enable measurements on more complex samples using sequential tilting 4D-STEM. In the future, combining sequential tilting 4D-STEM with other STEM-based electric characterization techniques, such as STEM EBIC \cite{meyer2021tracing} and in situ biasing, will enable reliable, quantitative, and comprehensive measurements of junction properties of real devices on the nanometer scale.
In addition, we expect that sequential tilting 4D-STEM will help to develop strategies to suppress dynamic diffraction effects in other applications like domain, phase, or strain mapping as well.

\section*{CRediT authorship contribution statement}
\textbf{Christoph Flathmann}: Conceptualization; Data curation; Formal analysis; Investigation; Methodology; Software; Validation; Visualization; Writing – original draft; Writing – review \& editing. \textbf{Ulrich Ross}: Investigation; Writing – review \& editing. \textbf{Jürgen Belz}: Resources; Writing – review \& editing. \textbf{Andreas Beyer}: Resources; Writing – review \& editing. \textbf{Kerstin Volz}: Funding acquisition; Resources; Writing – review \& editing. \textbf{Michael Seibt}: Funding acquisition; Supervision; Writing – original draft; Writing – review \& editing. \textbf{Tobias Meyer}: Conceptualization; Data curation; Methodology; Software; Project administration; Writing – original draft; Writing – review \& editing.

%\section*{Declaration of competing interest}
%The authors declare that they have no known competing financial interests or personal relationships that could have appeared to influence the work reported in this paper.

\section*{Acknowledgements}
This work was supported by the Deutsche Forschungsgemeinschaft (DFG, German Research Foundation) – 217133147/SFB 1073, projects B02 and Z02. The use of equipment of the “Collaborative Laboratory and User Facility for Electron Microscopy” (CLUE, Göttingen) is gratefully acknowledged.

%\section*{Data availability}
%The data used to determine the potential profiles across the \mbox{AlGaAs/GaAs} interface for different sample orientations is available from ... . Any other data that supports the findings of this study is available from the corresponding authors upon reasonable request.

%\section*{Code availability}
%The code to calibrate the microscope and to acquire the data as well as the code used for data analysis is available from https://gitlab.gwdg.de/christoph.flathmann/tilted-mrstem.

\bibliography{bibliography.bib}

\end{document}